\documentclass[
aps,%
12pt,%
final,%
notitlepage,%
oneside,%
onecolumn,%
nobibnotes,%
nofootinbib,%
superscriptaddress,%
noshowpacs,%
]%
{revtex4}
\usepackage{graphicx}

\begin{document}

\title{
Unveiling the nature of red novae cool explosions\\
using archive plate photometry
}
\author{\firstname{Vitaly}~\surname{Goranskij}}\email{goray@sai.msu.ru}
\author{\firstname{Natalia}~\surname{Metlova}}
\author{\firstname{Alla}~\surname{Zharova}}
\affiliation{Sternberg Astronomical Institute, Moscow University,
Russia}

\author{\firstname{Sergei}~\surname{Shugarov}}
\affiliation{Sternberg Astronomical Institute, Moscow University,
Russia}
\affiliation{Astronomical Institute, Slovak Academy of Sciences
}

\author{\firstname{Elena}~\surname{Barsukova}}
\affiliation{
Special Astrophysical Observatory, Russian Academy of Sciences
}

\author{\firstname{Peter}~\surname{Kroll}}
\affiliation{Sternwarte Sonneberg, Germany}

\begin{abstract}
\vspace{5mm}
Based on archive photographic photometry and recent CCD photometric
data for red novae V4332 Sgr and V838 Mon, we established their stellar
composition, exploded components, and the nature of explosions. Low
temperature in the outburst maximum is due to quasi-adiabatic expansion
of a massive stellar envelope after the central energy surge preceded
the outburst.
\end{abstract}

\maketitle


\vspace{10mm}

Red novae are the stars erupting into cool supergiants \cite{MHCZ02}.
There are images of progenitors for red novae V4332 Sgr and V838 Mon
in Moscow and Sonneberg plate collections. The first star is an
age-old object seen at the latitude of --9.4$^\circ$ and located in the
Galactic bulge or in the thick disk, the second one is a young
object associated with a cluster of B type stars and a dust
environment located at a distance of 6.1$\pm$0.6 kpc.
 
Three pairs of deep one-hour exposures in $B$ and $V$ bands of V4332 Sgr
were taken with the meniscus 50 cm Maksutov telescope AZT-5 of the
SAI Crimean Station between 1977 June and 1986 June. For V838 Mon
we have 148 $B$-band plates suitable for eye estimates, dated between
1928 and 1994 and taken with identical 40 cm astrographs in Sonneberg
and Crimea. 57 of them are good for digitization and accurate
measurements. In addition, there are 50 $V$-band plates obtained with
AZT-5, all are good for measurements. Emulsions of AGFA and ORWO
ZU-21 (Germany) were used to reproduce the $B$ band with the
40-cm astrographs, the ORWO ZU-21 was used with the BS-8 filter
at AZT-5 to cut the ultraviolet part of a spectrum. The Kodak 103aD
emulsion (USA) with the GS-17 filter was used to fit $V$ band at
the AZT-5. So, the observations were performed in the standard
bands of Johnson $UBV$ system.

To digitize plates, the Eastman Kodak CREO scanner of the Sternberg
Institute was used. Scanner output images in TIFF format were transformed
to BITMAP by MaxIm DL software with changing cuts. Additionally we used
the FinePix F10 FujiFilm camera in gray mode with an ordinary convex
lens, and transformed its JPEG images to BITMAP with MS Paint. The last
method of digitization can't be used for wide fields due to lens
distortion, but it is as good for a single star with outskirts as the CREO
scanner. Self-focusing and very short exposures of the FinePix
F10 camera allow shooting without support stands. So, this method may
be widely used by students and amateurs. Digital BITMAP images were
reduced with Goranskij's software WinPG, the characteristic curves were
plotted with 17 to 23 comparison stars and approximated with the 1st or
2nd order polynomial. The mean squire residuals equal to an average of
0$^m$.08 -- 0$^m$.12, but vary in the range of 0$^m$.04 -- 0$^m$.23
depending on the size of emulsion grains that is typical for photography.

The historical light curve of V4332 Sgr in the B band based on DSS,
AZT-5 plates, and modern CCD data is shown in Fig.~1. The outburst is not
plotted. The progenitor was a binary system consisting of a blue and
a red star. The $B$-band data along with other filters shows pre-outburst
brightening similar to that one detected in V1309 Sco by
OGLE \cite{THK11}. After the outburst, the brightness level fell down
below its level before the outburst
due to disappearance of the exploded blue companion. The second brightness
decay happened between 2006 and 2008, and was accompanied by the temperature
drop of the red M-type star by \hbox{1000 K.}

The historical light curve of V838 Mon in the $B$ band based on archival
plates, and on modern CCD data is given in Fig.~2. The outburst is not
presented in this Figure, too. Our photometry does not show any
significant variability of the progenitor and reveals that it was a system
of two B-type stars. It weakened by 0$^m$.461 $R$ in 1998, 4 years before the
outburst \cite{KE06}. In October 2002 the explosion remnant became so
cool (1200 K) that its radiation was not visible in $UBV$ bands, where
the light of the B3V star that survived after the explosion was dominating.
With the known magnitudes of the companion and the progenitor, we have
determined magnitudes of the exploded star. It was a B3V star, too, and
it was brighter than its companion by 36 per cent. In December 2006 the
explosion remnant evolved to an M-type giant. During the approach to the
remnant, the B3V companion disappeared totally for 70 days
in all the photometric bands,
what allowed us to measure its $UBVRI$ magnitudes. In 2007 we observed
a submergence of the companion into the remnant. During 200 days the hot
companion moved in the void under an exterior shell of the remnant, and
the radiation of the companion was absorbed by a factor of five. Then in
2008 it disappeared, and was not visible up to present time.

Finally, we were able to determine the spectral energy distributions
(SED) of progenitors and remnants of both red novae (Fig.~3). We found
that in V4332 Sgr SEDs (Fig.~3, left), the continuum of the M giant was
visible both before and after the outburst, and it was stronger and hotter
when it was closer to the outburst in time. We inclined to treat
the explosion in V4332 Sgr as a merger event in a blue straggler which
might be a contact binary star in a system with the M giant. We think that
continuum strengthening of the M giant was connected with the accretion
of matter onto its surface in the stages of forming common envelope of the
binary star and of a dynamical destruction of the merger remnant.

To extract SEDs of V838 Mon components from the common light of the binary
system,
we measured the light of each component lost in the eclipse or after the
explosion. The SED of the exploded star (central one of the three in
Fig.~3, right) was determined as a difference of the progenitor's SED
and the survived B3V companion's SED. The SED of the exploded star is
compared with the SED of HD 29763 (B3V). With the known distance, it
is found to be located in the zero-age main sequence of the Temperature --
Luminosity diagram. The exploded component of the system was a young star
with R = 2.9 R$_\odot$, L =1020 L$_\odot$, log T$_{eff}$ = 4.29. Its companion
is a lower luminosity star having R = 2.5 R$_\odot$, L =740 L$_\odot$ and
the same temperature. There is no evidence of binary nature or merger for
the exploded star in V838 Mon. In addition, we established that the radius
of the remnant at the first appearance in the outburst with K0 I spectrum
was equal to 327 R$_\odot$, and that the exploded star envelope had
undergone pre-outburst expansion in the conditions close to adiabatic
which continued at least four years. The central energy surge causing
a slow shock to massive star envelope is a reason of cool explosions
of red novae \cite{BGVZ14}.

In a massive star experienced such an energy surge, the radiation transfer
time exceeds its dynamic expansion time by many orders of value, so the
explosion energy is concentrated in the bottom of the expanding envelope.
The surface area of the envelope becomes very large when the radiation
reaches it, and the explosion energy is insufficient to heat the star
surface to a high temperature. Reasons of such energy surges may be both
the merger of stellar nuclei after forming a massive common envelope
in a contact binary, and the instability in the nucleus of a massive young
star. So, the red nova phenomenon is representative of both old and young
stellar populations.\\

\begin{figure*}
\includegraphics[scale=0.48]{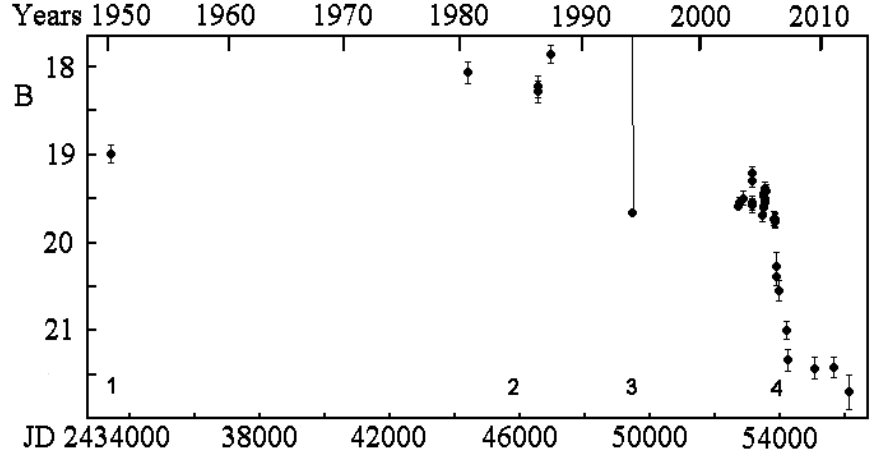}
\caption{
Historical light curve of V4332 Sgr in the $B$ band. The main points of the
evolution are marked. 1-- POSS~I. 2 -- AZT-5 + POSS II; brightening before
the outburst. 3 -- The outburst. 4 -- The light and temperature decay of
the M-type star after the outburst.
}
\end{figure*}

\begin{figure*}
\includegraphics[scale=0.48]{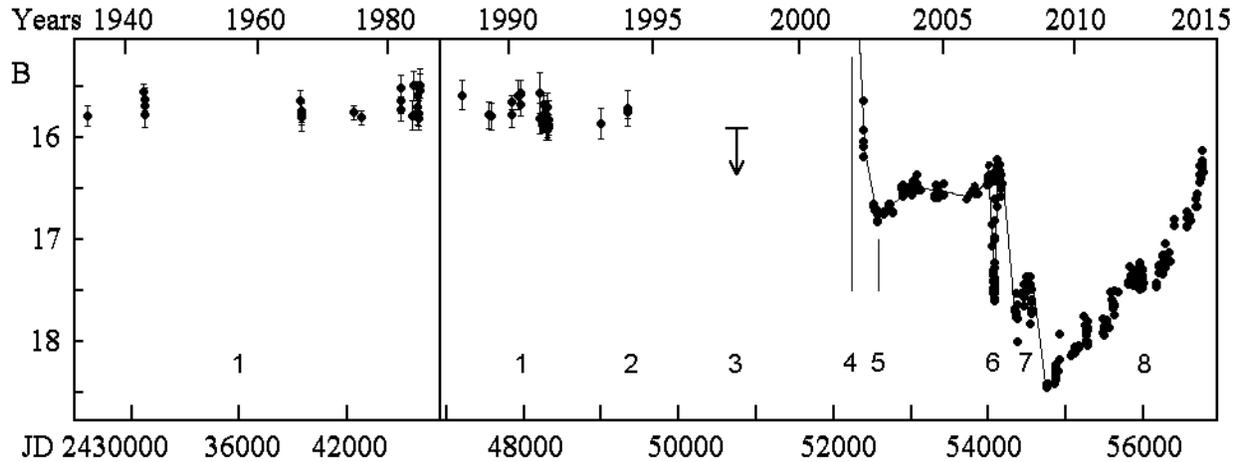}
\caption{
Historical light curve of V838 Mon in the $B$ band. Main phases of the
evolution are noted. 1 -- Progenitor, the pair of B3V stars. 2 -- The
astro-plate production stopped. 3 -- Kimeswenger \& Eires found a decay in
the $R$ band. 4 -- Outburst of the brighter B3V star in the binary.
5 -- B3V companion is only visible in the $B$ band. 6 -- The eclipse.
The B3V companion vanished for 70 days. 7 -- The B3V companion moving in
the void is visible through the shell. 8 -- The B3V companion plunged inside
the M-type remnant. The remnant becomes brighter.
}
\end{figure*}

\begin{figure*}
\includegraphics[scale=0.40]{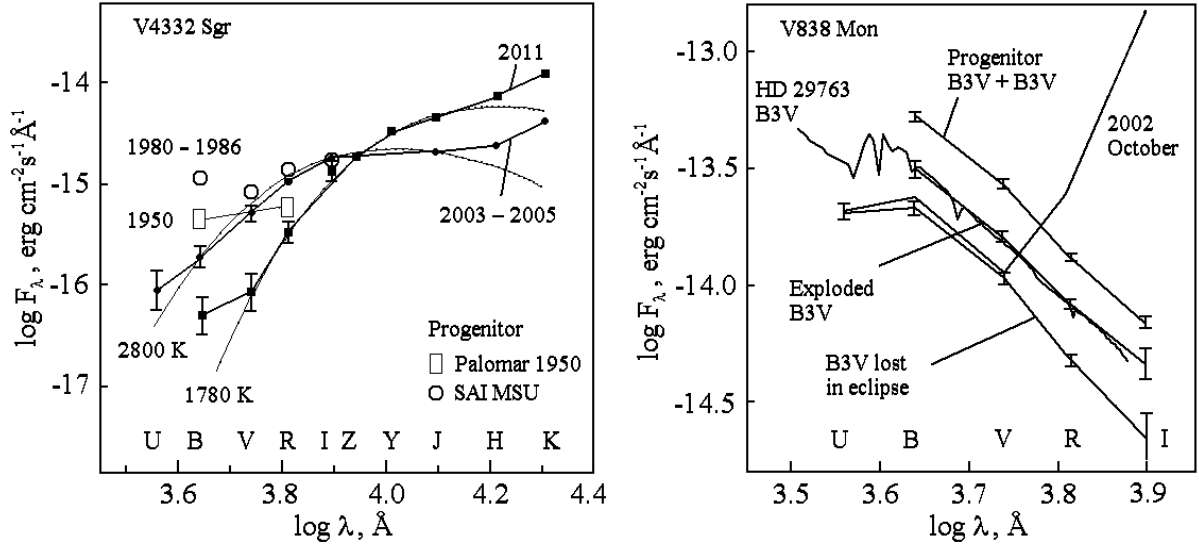}
\caption{
Spectral energy distributions (SED) of V4332 Sgr (left) and V838 Mon (right).
All SEDs are corrected for interstellar reddening. Empty signs in V4332 Sgr
SED are photographic observations of the progenitor, black signs and lines
mark M-type star continuum without emission-line contribution. Blackbody
fitting is also shown. SEDs extracted from the common light of V838 Mon
(right). The following SEDs are given: the progenitor binary (top), the
exploded component before its explosion compared with the SED of HD 29763
(B3V) (middle), and the B3V companion (bottom). The two-component SED of
the binary with the cool remnant in 2002 October is also shown.
}
\end{figure*}

\end{document}